\documentclass[12pt,preprint]{aastex}

\slugcomment{Draft, \today}

\begin{document}

\title{Molecular CO outflows in the L1641-N cluster: kneading a cloud core}


\author{T. Stanke\altaffilmark{1,2}}
\affil{European Southern Observatory, Karl-Schwarzschild-Str.\ 2, 85748 Garching, Germany}

\author{J. P. Williams}
\affil{Institute for Astronomy, University of Hawai'i,
    Honolulu, HI 96822}




\altaffiltext{1}{Institute for Astronomy, University of Hawai'i,
    Honolulu, HI 96822}
\altaffiltext{2}{Max-Planck-Institut f\"ur Radioastronomie
Bonn, Auf dem H\"ugel 69, 53121 Bonn, Germany}

\begin{abstract}
We present results of 1.3~mm interferometric and single-dish observations
of the center of the L1641-N cluster in Orion. Single-dish wide-field
continuum and CO(2-1) observations reveal the presence of several molecular
outflows driven by deeply embedded protostellar sources. At higher angular
resolution, the dominant millimeter source in the cluster center is resolved
into a pair of protostars (L1641-N-MM1 and MM3), each driving a collimated
outflow, and a more extended, clumpy core. Low-velocity
CO line-wing emission is widely spread over much of the cluster area.
We detect and map the distribution of several other molecular transitions
($^{13}$CO, C$^{18}$O, $^{13}$CS, SO, CH$_3$OH, CH$_3$CN, and OCS).
CH$_3$CN and OCS may indicate the presence of a hot corino around L1641-N-MM1.
We tentatively identify a velocity gradient over L1641-N-MM1 in CH$_3$CN and
OCS, oriented roughly perpendicular to the outflow direction, perhaps indicative
of a circumstellar disk. An analysis of the energy and momentum load of the
CO outflows, along with the notion that apparently a large volume fraction
is affected by the multiple outflow activity, suggests that outflows from a
population of low-mass stars might have a significant impact on clustered
(and potentially high-mass) star formation.
\end{abstract}



\section{Introduction}

Most stars are born in clustered environments \citep[e.g.,][]{ladalada2003}
and their formation may be significantly different than in isolated
regions such as the well studied Taurus cloud.
For example, the shape and extent of prestellar cores and their subsequent
evolution may be affected through close encounters with other cluster
members or the effect of ionizing radiation from massive cluster stars.
Protostellar outflows, known to accompany star formation, might also play
a different role in clusters. Whereas in the isolated case much of the
outflow energy and momentum will likely just be dumped somewhere else in or
even outside the cloud, multiple, randomly oriented outflows in a 
cluster-forming clump might impact a larger volume, significantly affecting
or even hindering further star formation.

We present new data on the outflow activity in the L1641-N region in the
Orion~A giant molecular cloud. As a moderately sized, relatively
nearby\footnotemark
\footnotetext{we assume the canonical Orion distance of 450~pc}
cluster, L1641-N allows for detailed observations, in a relatively simple
example, of the complex processes involved in cluster formation.

The first detection of bipolar CO outflow activity in L1641-N,
centered on IRAS~05338$-$0624, was found by \cite{fukuietal1986}
and subsequently followed up by \cite{fukuietal1988} and \cite{wilkingetal1990}.
Near infrared imaging revealed the presence of an embedded small group
of about 20-25 young stars \citep{strometal1989,chenetal1993,hodappdeane1993}. 
\cite{hodappdeane1993} estimated that the cluster grew at a rate of
2-3 stars every 10$^5$~yr over about 10$^6$~yr.

The CO outflow maps have generally been interpreted as evidence for the
presence of one, roughly north-south oriented outflow, driven by one 
protostellar source seen as the mid- to far-infrared source IRAS~05338$-$0624.
Ground based $M$-band observations have revealed a source N1 seen
at 5~$\mu$m (and longward) only \citep{chenetal1993};
this source is also seen as
a compact millimeter continuum source in interferometric images 
\citep{wilkingetal1989,mcmullinetal1994,chenetal1995}, but appears not to
be the dominant source at least at mid-infrared wavelengths 
\citep[][AN04 in the following]{stankeetal1998,alinoriegacrespo2004}. 
Moreover, interferometric CS and SiO observations
\citep{mcmullinetal1994} suggest a south-west to north-east oriented
outflow at the very cluster center, as opposed to the predominantly
north-south orientation seen on larger scales in CO. \cite{stankeetal1998}
have proposed that there are indeed 2 independent outflows: a south-west to
north-east oriented flow, and a large-scale,
north-south outflow, driven by the brightest 10~$\mu$m source in the field.
The south-west to north-east flow appears to be driven by the compact
millimeter continuum source.
Further optical and near-infrared narrow-band imaging
searches for Herbig-Haro objects and shocked H$_2$ emission have provided
evidence that the L1641-N cluster indeed is the center of multiple outflows 
\citep[][(hereafter SMZ2002)]{davisetal1995,reipurthetal1998,stankeetal1998,maderetal1999,stankeetal2000,stankeetal2002},
some of which extend over several parsecs.

In this contribution we present new 1.3~mm wavelength single-dish and
interferometric observations of the L1641-N cluster area in order
to shed further light on the protostellar outflow activity. We will first
describe the observations (Section \ref{chap_obs}), identify outflows and
their driving sources in Section \ref{chap_results}, and provide a
summary and conclusions in Section \ref{chap_summary}.

\section{Observations \& data reduction
\label{chap_obs}}

\subsection{IRAM 30~m}

We obtained 1.3~mm dust continuum maps in the course of a wider survey over
several observing runs between 1999 and 2002, using the 37 channel
MAMBO bolometer array at the IRAM~30m telescope. The data reduction was done
with MOPSI\footnote{developed and maintained by Robert Zylka, IRAM
Grenoble, France} and followed standard bolometer
reduction procedures, including baseline subtractions, despiking, correction
for athmospheric extinction using the results of skydip measurements,
and flux calibration using planet maps.
Correlated sky-brightness variations (sky-noise) were removed in an iterative
process, using the resulting mosaic of the respective previous iteration as
input source model to the actual sky-noise reduction. Thereafter, the
single beam intensity distribution was restored from the chopped dual-beam
data and subsequently combined into the large mosaic.

The 9 element heterodyne receiver array HERA \citep{schusteretal2004}
was used in November 2003
to obtain CO(2-1) maps of the L1641-N region. The cluster center and
the area around the driving source of Outflow~51 in SMZ2002 
were each covered with a fully sampled, 
66\arcsec{}$\times$66\arcsec{} raster
map. In Addition, a $\sim$10\arcmin{}$\times$9\farcm5 area was mapped in the
coarsely sampled spectral line On-The-Fly 
mode\footnote{sampling at 7.6\arcsec{}; see HERA manual
for details: http://www.iram.es/IRAMES/mainWiki/HeraWebPage}.
Data reduction included low-order baseline
subtraction and combining the spectra (raster and OTF maps) into one map.

To convert CO brightness temperatures to molecular gas column densities,
we assumed the CO to be optically thin, in LTE at a temperature of 30~K,
and a CO abundance of $10^{-4}$ of the H$_2$ abundance. We derived gas masses
as a function of velocity for each channel. Kinetic energies and momenta were
derived as $E_{\rm kin} = 1/2 \sum M(v)\cdot (v - v_{\rm cen})^2$ and
$P = \sum M(v)\cdot (v - v_{\rm cen})$, with $v_{\rm cen}=6.8$~km/s.
A characteristic timescale was derived
by dividing the maximum distance $l$ at which high-velocity emission was seen
by the maximum CO velocity observed. Results for the various outflow lobes
identified from the data are listed in Tab.\ \ref{tab_COlobes}.

We do neither attempt to correct for optical depth effects in calculating
the CO outflow masses nor for outflow gas at radial velocities equal to the
ambient cloud velocities; similarly, we do not correct the velocity of the
gas for projection effects. Outflow masses, kinetic energies, and momenta
are therefore strict lower limits.

\subsection{SMA}

We used the Submillimeter Array (SMA) on Mauna Kea, Hawai'i, to obtain
high resolution 230~GHz images of the center of the L1641-N
cluster. Two tracks were taken, one in the compact (Dec.\ 19, 2004,
with antennas 1--6; antenna 8 was flagged)
and one in the extended configuration (Dec.\ 27, 2004, with antennas
1--6 and 8). The two configurations yielded projected baselines ranging
from about 13 to 220~m.
Double sideband receivers were used, with the CO(2-1)
line centered in the upper sideband (correlator chunk 14). 230~GHz
zenith opacities were between 0.1 and 0.06 during the first track, with
fairly stable phases; system temperatures were between 200 and 250~K throughout
most of the track. During the second track, the opacity was initially high,
$\sim$0.2, but dropped to $\sim$0.08 later on; similarly, phase
stability improved during the track. System temperatures were around 150~K.

Data calibration and editing were done within the IDL MIR package. 
The phase center for L1641-N was set to 
$\alpha$=5$^{\rm h}$36$^{\rm m}$18.$^{\rm s}$8, 
$\delta$=-6$^\circ$22\arcmin10\arcsec (J2000).
Observations of Uranus and Callisto were used for
bandpass and flux calibration. Flux calibration was obtained using
model predictions for Callisto's flux depending on baseline length
as provided on the SMA webpages\footnotemark
\footnotetext{http://sma1.sma.hawaii.edu/}.
Observations of J0423$-$013 and J0607$-$157 were taken every 20~min to
allow for complex phase calibration.
Continuum emission was reconstructed from line-free parts of the 
2$\times$2~GHz range covered by the SMA receivers.

Maps were constructed and deconvolved within MIRIAD. For the emission line
maps, continuum emission was first subtracted from the emission line
visibilities. The continuum and $^{12}$CO line SMA data were complemented
with short spacing visibilities obtained from the IRAM 30~m mapping
described above. Maps were created using a natural weighting scheme,
and CLEANing was performed, stopping after new CLEAN components were at
about 60\% of the rms noise level.
The synthesized beam measures 1\farcs7$\times$1\farcs2 at a position angle
of 75$^\circ$ east of north.

\subsection{Astrometry}

In this paper we use a number of different data sets, taken over a wide
wavelength range: near-infrared ($K$-band) data from 
SMZ2002, ISO data published by AN04,
and new single-dish and interferometric millimeter wavelength data. 
Good relative astrometry is crucial in understanding what happens in a 
clustered environment such as L1641-N. The various datasets used in this
paper were registered with respect to each other as follows: the infrared
$K$-band data from SMZ2002 were recalibrated over the
field of interest using 2MASS coordinates for stellar sources. The ISO maps
were registered to the near-infrared using a number of sources seen in the
near- and mid-infrared maps. The 1.2~mm dust continuum map was registered
with respect to the near-infrared map using one star seen in both maps, 
located 1\farcm5 south-east of the cluster center. The interferometer map
was not corrected, because it is phase referenced to quasars, yielding high
positional accuracy.

\section{Young stellar objects and outflows in L1641-N
\label{chap_results}}

\begin{figure}[th]
  \includegraphics[width=16cm]{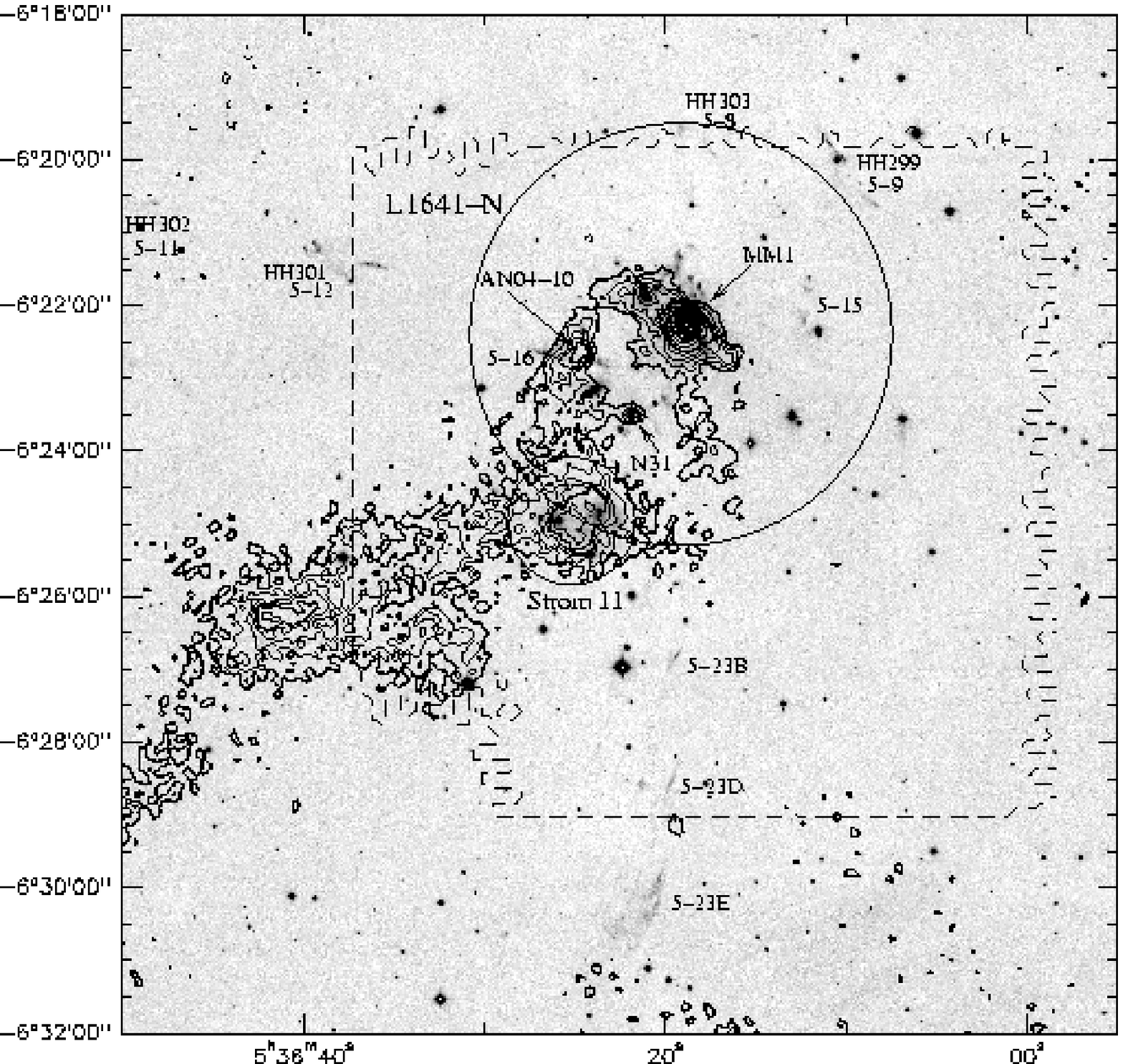}
  \caption{Overview of the area around the L1641-N cluster.
  Solid black contours represent 1.3~mm dust continuum (IRAM~30~m/MAMBO single-dish).
  Greyscale: 2.12~$\mu$m narrow band image, not continuum subtracted. HH objects and
  SMZ2002 shock excited H$_2$ features are marked (e.g., 5-12 is [SMZ2002]5-12).
 The dashed line marks the boundary of the
  field covered by the wide-field IRAM~30~m/HERA CO map.
  AN04-10 marks the 1.3~mm counterpart to the AN04 ISO mid-infrared source 10.
  N31 is source N31 of \cite{chenetal1993}.
  }
\label{fig_cont_nir}
\end{figure}

\begin{figure}[th]
  \includegraphics[width=16cm]{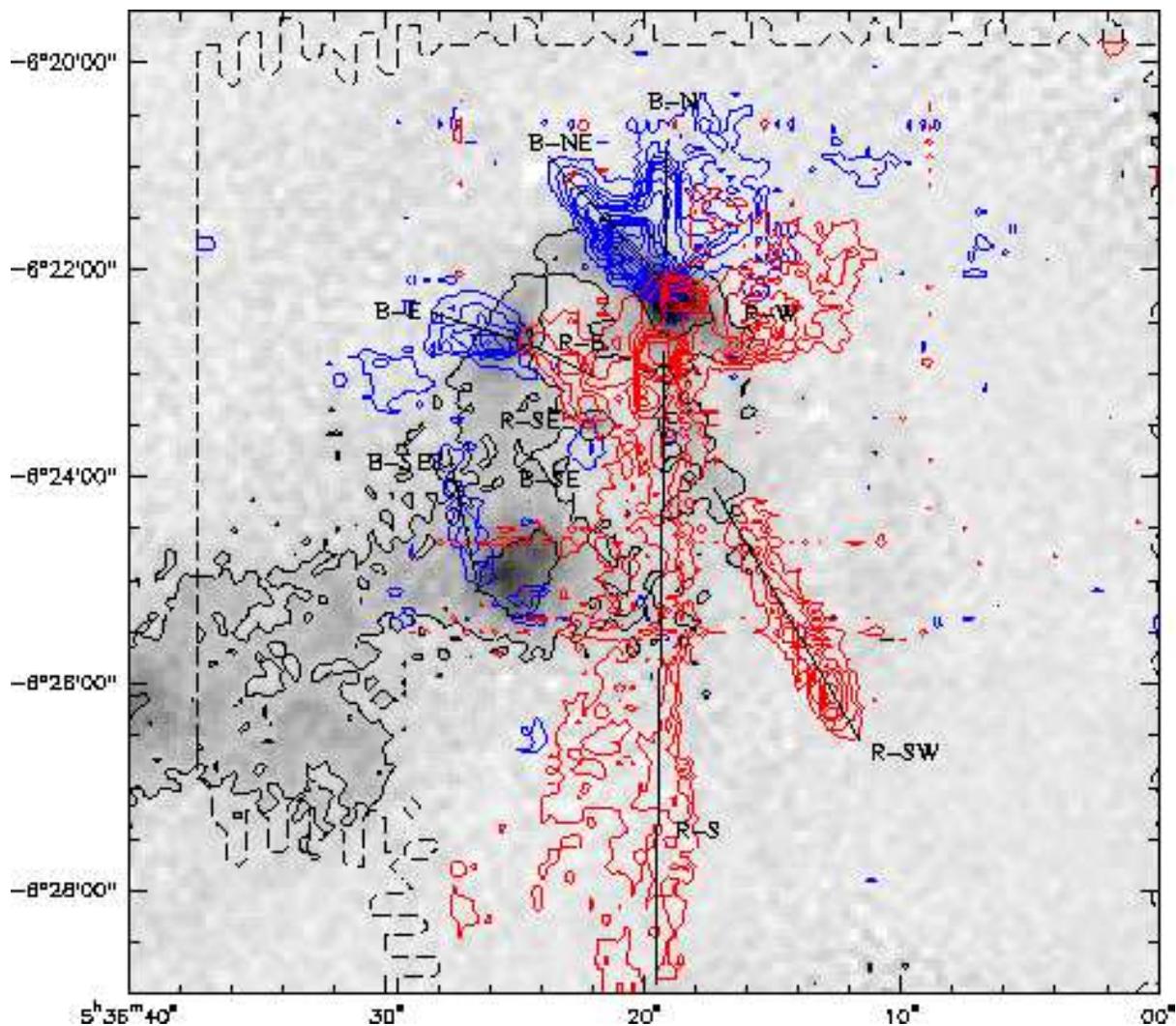}
  \caption{CO emission in the L1641-N cluster area. Blue/red countours:
  blue-/red-shifted CO (IRAM~30~m/HERA). Solid black
  contours \& greyscale: 1.3~mm dust continuum (IRAM~30~m/MAMBO single-dish).
  We suggest that B-E, R-E, and R-SW are part of one bipolar outflow driven
  by the AN04 ISO source 10, with the redshifted lobe
  being deflected by interaction with the L1641-N giant outflow (R-S). A small
  bipolar flow (R-SE, B-SE) is driven by the N31 source of  \cite{chenetal1993}.
  R-S and B-N are part of the L1641-N giant outflow. B-NE and R-W are resolved
  into two parallel outflows by our SMA observations (see Fig.\ \ref{fig_CO_SMA}). 
  }
\label{fig_COwide}
\end{figure}

Figure \ref{fig_cont_nir} shows an overview of the field around the L1641-N
cluster
in 1.3~mm dust continuum (contours) and 2.12~$\mu$m narrow band (greyscale:
H$_2$ $v$=1-0~S(1) line $+$ continuum; see \cite{stankeetal1998}, SMZ2002).
Fig.\ \ref{fig_COwide} shows the CO~$J=$2-1 line-wing emission 
(blue/red contours).

\subsection{Wide-field mapping}

The wide-field 1.3~mm dust continuum map taken with MAMBO shows a number
of compact features in the center of the field and more diffuse emission
towards the south-east (Fig.\ref{fig_cont_nir}).
This is reminiscint of the head-tail shape of the C$^{18}$O
cloud core as mapped by \citet{reipurthetal1998} (their Fig.\ 17a)
\citep[see also][]{tatematsuetal1993}. 

\begin{figure}
  \includegraphics[scale=1.03,angle=270]{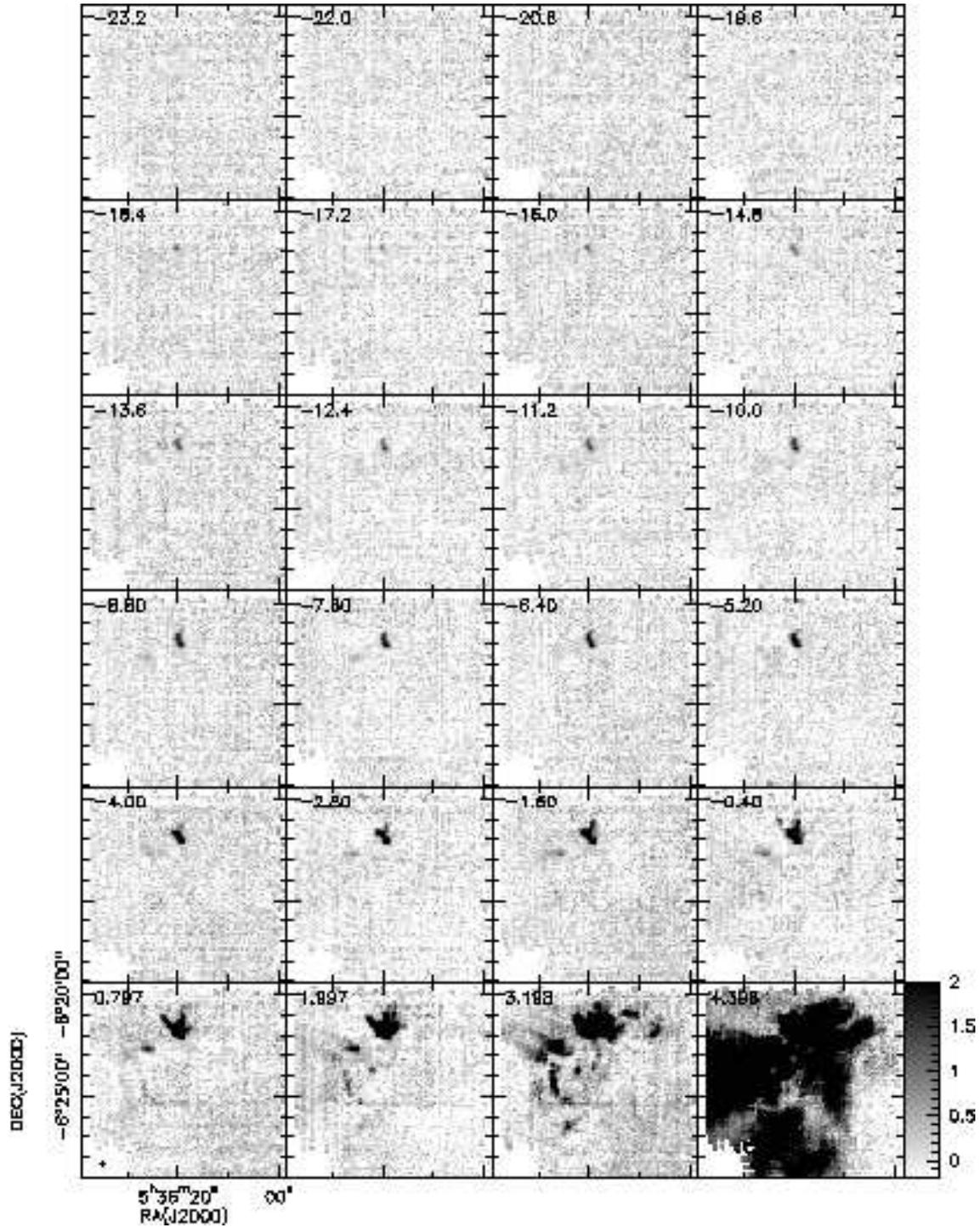}
  \caption{CO emission in the L1641-N cluster area: Channel maps of the
  blueshifted emission with velocities between -23.2 and 4.4\,km\,s$^{-1}$.
  Individual maps
  are labeled with the channels central velocity. The small circle in the lower
  left corner of the lower left panel indicates the beamsize (11\arcsec).
  }
\label{fig_COwide_chamaps_a}
\end{figure}

\begin{figure}
  \includegraphics[scale=1.03,angle=270]{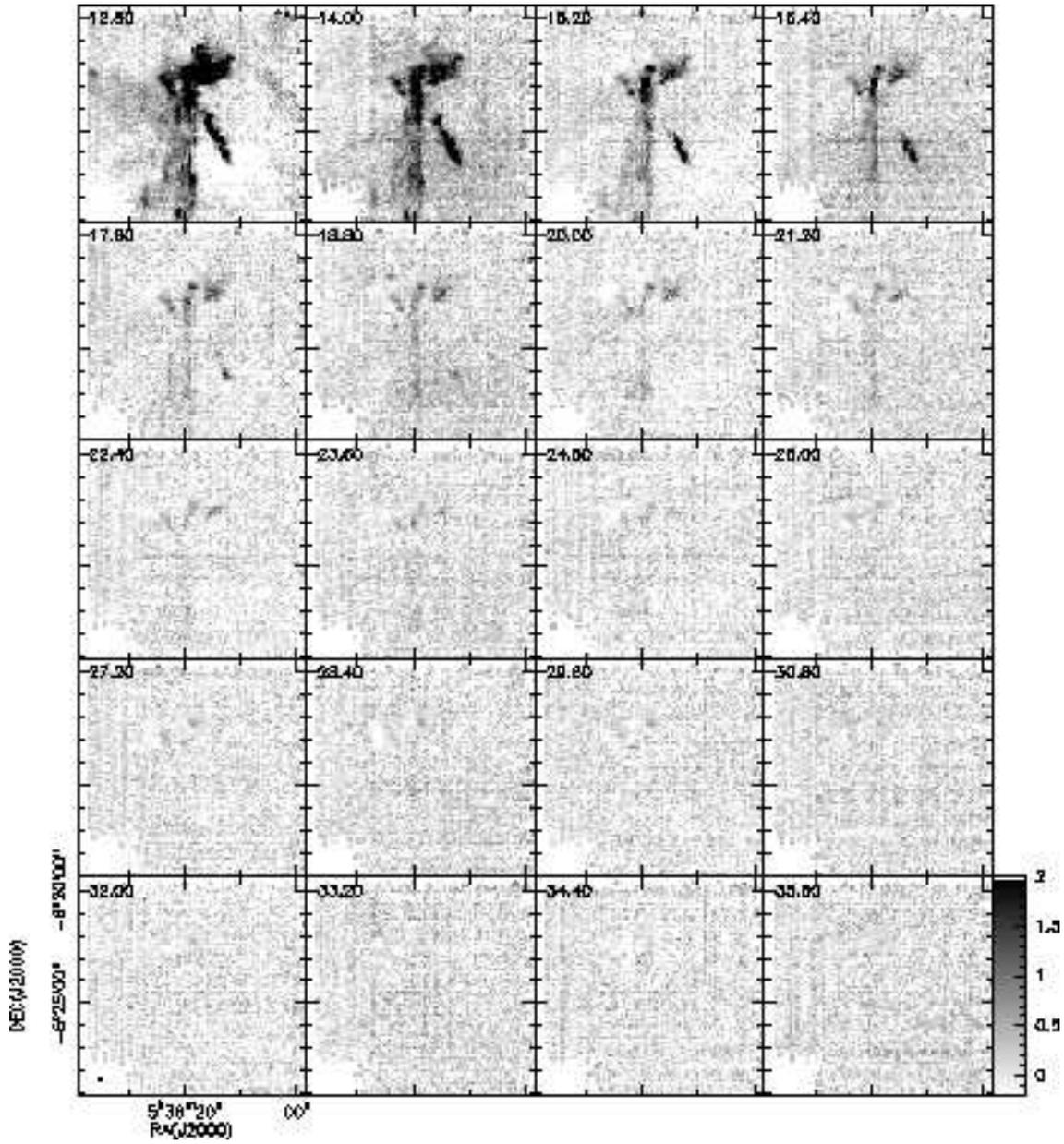}
  \caption{As Fig.\ \ref{fig_COwide_chamaps_a}, for redshifted emission
  (12.8 to 35.6\,km\,s$^{-1}$).
  }
\label{fig_COwide_chamaps_c}
\end{figure}

The CO line-wing emission maps (Figs.\ \ref{fig_COwide_chamaps_a} and
\ref{fig_COwide_chamaps_c}) reveal the presence of a number of
well-defined outflow lobes.
The dominant feature at red-shifted velocities is a $\sim$1\farcm5 broad,
collimated lobe extending over 7\arcmin{} south of the L1641-N cluster center
to the edge of the area mapped in CO (labeled R-S in Fig.\ \ref{fig_COwide}).
It is also seen as a series of H$_2$
shocks reaching even further south \citep{stankeetal2000}. Together
with a chain of optical Herbig-Haro objects north of the cluster 
\citep[HH~306-HH~310;][]{reipurthetal1998,maderetal1999}
these features form a bipolar giant outflow stretching over several parsecs
on both sides of the cluster. Blue-shifted CO emission is also seen north
of the cluster center in a 'Y'-shaped feature. The channel maps in Fig.\
\ref{fig_COwide_chamaps_a} show that at high velocities there is a collimated
lobe extending NE of the cluster center (B-NE), which is joined by an
apparently also collimated N-S feature (B-N) at intermediate
velocities. At low velocities, these features blend with a broad patch
of CO emission north of the cluster center.
Broadly distributed red-shifted emission is also seen to the
south-west and west of the cluster center (Fig.\ \ref{fig_COwide_chamaps_c}).
High-velocity emission
seems to be restricted to a small lobe close to the cluster center,
forming a counter-lobe to B-NE. The MAMBO dust continuum map shows
a bright, compact but resolved peak at the center of these CO outflow lobes,
on top of a north-east to south-west oriented ridge, which corresponds
to the HCN and HCO$^+$ ridge found by \cite{fukuietal1988}.
 
\begin{figure}
  \includegraphics[scale=0.8,angle=270]{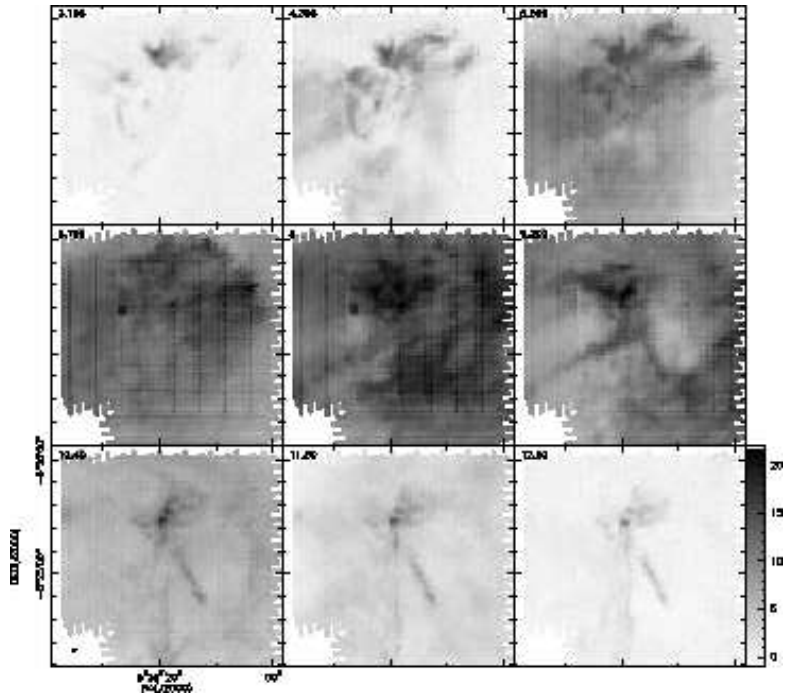}
  \caption{As Fig.\ \ref{fig_COwide_chamaps_a}, for velocities around the 
  ambient cloud velocity (3.2 to 12.8\,km\,s$^{-1}$, note the different
  greyscale stretch).
  }
\label{fig_COwide_chamaps_b}
\end{figure}

Another well defined, highly collimated outflow lobe (R-SW) is seen
to the west of the large N-S lobe R-S at a position angle of about
28$^\circ$ east of north.
The CO channel maps (Fig.\ \ref{fig_COwide_chamaps_c}) reveal a distinct
velocity structure in this lobe, with
the highest velocity emission being found at the south-western tip, and lower
velocity emission extending ever further to the north-east. In fact, close to
the ambient CO velocity (Fig.\ \ref{fig_COwide_chamaps_b}) this lobe structure
can be clearly followed up to 
where it intersects with the large, redshifted N-S giant flow lobe.
The MAMBO continuum map also shows a faint ridge along the northern portion
of this CO lobe. We suggest this is outflow entrained dust
\citep[c.f.,][]{chinietal2001}.
Converting the CO flux densities to broad-band flux densities yields
flux densities of the order of 2-5~mJy/beam (using a conversion factor
of 8.6~Jy/K as given on the IRAM webpages and a MAMBO bandwidth of 100~GHz
estimated from the transmission plots shown on the MPIfR MAMBO webpages),
much less than the 30-50~mJy/beam seen in the MAMBO map in the ridge. 
The R-SW lobe does not have any obvious blue-shifted counterlobe.
No H$_2$ shocks have
been found that could be associated with this lobe by SMZ2002.

A bipolar CO outflow is centered on 
$\alpha$=5$^{\rm h}$36$^{\rm m}$24.$^{\rm s}$8, 
$\delta$=-6$^\circ$22\arcmin42\arcsec. It
corresponds to [SMZ2002]~Outflow~51. The outflow is
centered on a north-west to south-east oriented millimeter continuum ridge
and an ISO mid-infrared source (source 10 of AN04).
Its blueshifted
lobe is seen over a large distance from the source as a sequence of large
optical HH-objects (HH~301 and HH~302) and infrared shocks 
\citep{reipurthetal1998,maderetal1999,stankeetal2002}. 
No such features have been found at comparable distances from the source
on the redshifted side. We speculate that the flow might
have a continuation on the redshifted side in the outflow lobe R-SW, for which
we cannot see any other plausible driving source. This would imply a 
significant change in outflow direction, however, and cannot be due to
precession as only the redshifted lobe shows this bend.
We noted that at velocities close to the CO ambient cloud 
velocity the R-SW lobe can be traced back to where it intersects with
the R-S lobe. This is also about where the R-E lobe of [SMZ2002]~Outflow~51
would intersect with R-S. We therefore speculate that Outflow~51
interacts with R-S, its outflowing gas being accelerated in a southerly
direction, thus creating the bend in the redshifted flow lobe.
Deflection of a jet with a side wind is a candidate mechanism to explain
bent jets \citep[e.g.,][]{fendtzinnecker1998}, particularly in the case of
'C'-shaped symmetry about the driving source \citep[e.g.,][]{ballyetal2006}.
Observations of jets of this type, as well as theoretical and experimental
studies, demonstrate
that jets can survive (and remain collimated) while interacting with relatively
tenous winds such as expanding HII regions 
\citep[e.g.,][]{masciadriraga2001,lebedevetal2004}
over a large portion of the jet length. 
In L1641-N in contrast, the R-SW lobe seems to interact with R-S only over a
short total length, and it seems that R-S is heavier than R-SW, as R-S does
not show any signs of bending resulting from an interaction with R-SW.
That a lighter jet might be able to survive a collision with a heavier flow
remains to be shown, but it seems that a collision of two jets and redirection
does not necessarily imply a disruption of the flow(s) 
\citep{cunninghametal2006}.

In addition, we identify
a small, bipolar CO outflow centered on a near- to mid-IR
\citep[source N31 of][ISO source 13 of AN04]{chenetal1993}
and 1.3~mm continuum source
at $\alpha$=5$^{\rm h}$36$^{\rm m}$22.$^{\rm s}$0,
 $\delta$=-6$^\circ$23\arcmin28\arcsec. The blueshifted lobe is
associated with two faint H$_2$ shocks which were not identified by SMZ2002.

Finally, the Strom~11 group of embedded stars
\citep{strometal1989b,chenetal1993} is seen superimposed on a
roughly circular millimeter continuum clump of $\sim$1\arcmin{} diameter.
A faint, narrow lobe of blueshifted CO emission (B-SE2, seen best in the
channel maps in Fig.\ \ref{fig_COwide_chamaps_a}, close to the ambient velocity)
might have its origin in the Strom~11 group; some more faint, blueshifted 
emission is seen in Strom~11 and to its south.

\subsection{The central area: SMA observations}

\subsubsection{Continuum}

\begin{figure}[th]
  \includegraphics[width=16cm]{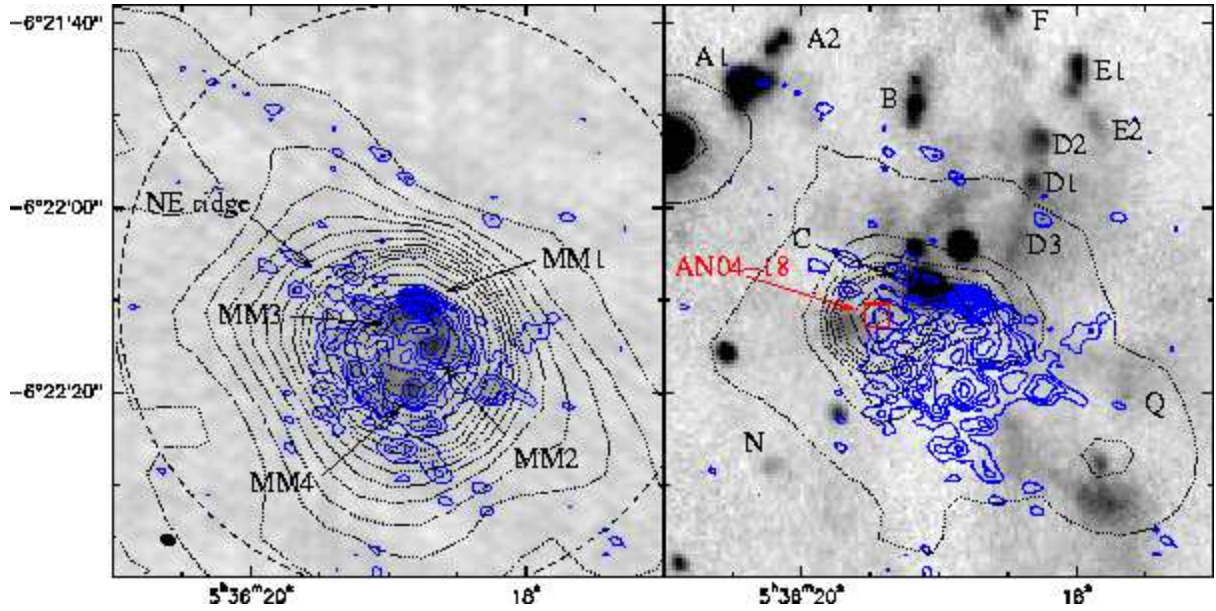}
  \caption{1.3~mm dust continuum emission in L1641-N. Left panel: SMA+MAMBO
  combined map (greyscale, blue contours), with the single-dish MAMBO map
  overlayed
  as dotted contours. The SMA beamsize is indicated as the black ellipse, and
  the 30\% sensitivity radius of the SMA primary beam is shown as the dashed
  circle. Right: the greyscale shows a 2.12~$\mu$m near-infrared narrow band
  image ($K$-band continuum + H$_2$ $v$=1-0~S(1) line, see 
  \cite{stankeetal2002}), with H$_2$ shocks labeled; dotted contours:
  ISO 7~$\mu$m map (AN04). The square marks the position of AN04 ISO 
  source 18.} 
\label{fig_continuum}
\end{figure}

Figure \ref{fig_continuum} shows the 1.3~mm dust continuum at high angular
resolution (SMA and MAMBO data combined). 
We detect a bright, compact source centered at
$\alpha =$5$^{\rm h}$36$^{\rm m}$18.$^{\rm s}$780, 
$\delta =-$6$^\circ$22\arcmin10\farcs37 (J2000), in very good agreement
with the position found by \cite{chenetal1995}. We will
refer to this source as L1641-N~MM1 in the following. 
The visibilities decrease from 0.3-0.4~Jy at short spacings to
0.1-0.2~Jy at large uv-distances, so the source is resolved.
A gaussian fit to the radially averaged real continuum
visibilities implies a FWHM size of $1\farcs 0\pm 0.1$.
From elliptical gaussian fits in the image plane we find a size of
2\farcs1$\times$1\farcs5 (P.A. 67$^\circ$), corresponding to
a deconvolved size 1\farcs2$\times$0\farcs8 (P.A. 56$^\circ$). 
Fitting the combined MAMBO+SMA data at
fluxes $>$0.02~Jy, gives a size 2\farcs5$\times$1\farcs7 (P.A. 67$^\circ$),
corresponding to 1\farcs9$\times$1\farcs3 (P.A. 62$^\circ$) deconvolved.
The image shows a tail
of emission running at a similar position angle away from the source,
so the P.A. derived from the gaussian fit might be biased towards this
direction. As the deconvolved source size is significantly smaller than
the beam size, the result is likely subject to large errors.
Still it seems that the continuum emission is not oriented perpendicular to the
outflow direction inferred from the CO observations (see below).
As the emission of MM1 furthermore blends into more extended core emission,
we suggest that MM1 traces an envelope structure rather than a disk.
The total integrated flux of 0.54~Jy (as measured from the gaussian fit)
corresponds to a circumstellar gas mass of about 1.6~$M_\odot$,
assuming a dust temperature of 20~K and a dust
opacity coefficient of 0.01~cm$^2$g$^{-1}$ \citep{motteetal1998}.

To the south of MM1
a number of extended features form a clumpy core. The most prominent of these
are two clumps to the south of L1641-N~MM1 (L1641-N~MM2 and L1641-N~MM4 in
the following). There appears
to be an extension of L1641-N~MM1 in a SW direction, forming a bridge of 
emission between L1641-N~MM1 and L1641-N~MM2. Another extension to L1641-N~MM1
is seen to protrude towards the ESE, which appears as a separate source in
some of the observed molecules (see below); we will refer to this feature as
L1641-N~MM3.

Protruding to the NE of L1641-N~MM3 there appears to be a clumpy ridge of
emission parallel to the small H$_2$ jet (feature C in Fig.\ 
\ref{fig_continuum}). This ridge might outline the wall of a cavity produced
by the jet.
Finally, some more clumpy emission is seen to the south and west. Generally,
there is a good anticorrelation between the millimeter continuum emission and
the diffuse infrared emission, indicating that the dust is in the foreground
of the diffuse emission.

Notably, the single-dish continuum peak does not coincide with the position
of MM1; although pointing errors might affect the single-dish peak position,
the offset seems to be too large; note also that the astrometry of the
single-dish map has been corrected using a near-infrared source, the position
of which should be accurate to better than 1~arcsecond. Thus the single-dish
continuum peak seems to trace the clumpy core south of MM1 rather than MM1
itself.

\subsubsection{$^{12}$CO line-wing emission}

\begin{figure}
  \includegraphics[width=16cm]{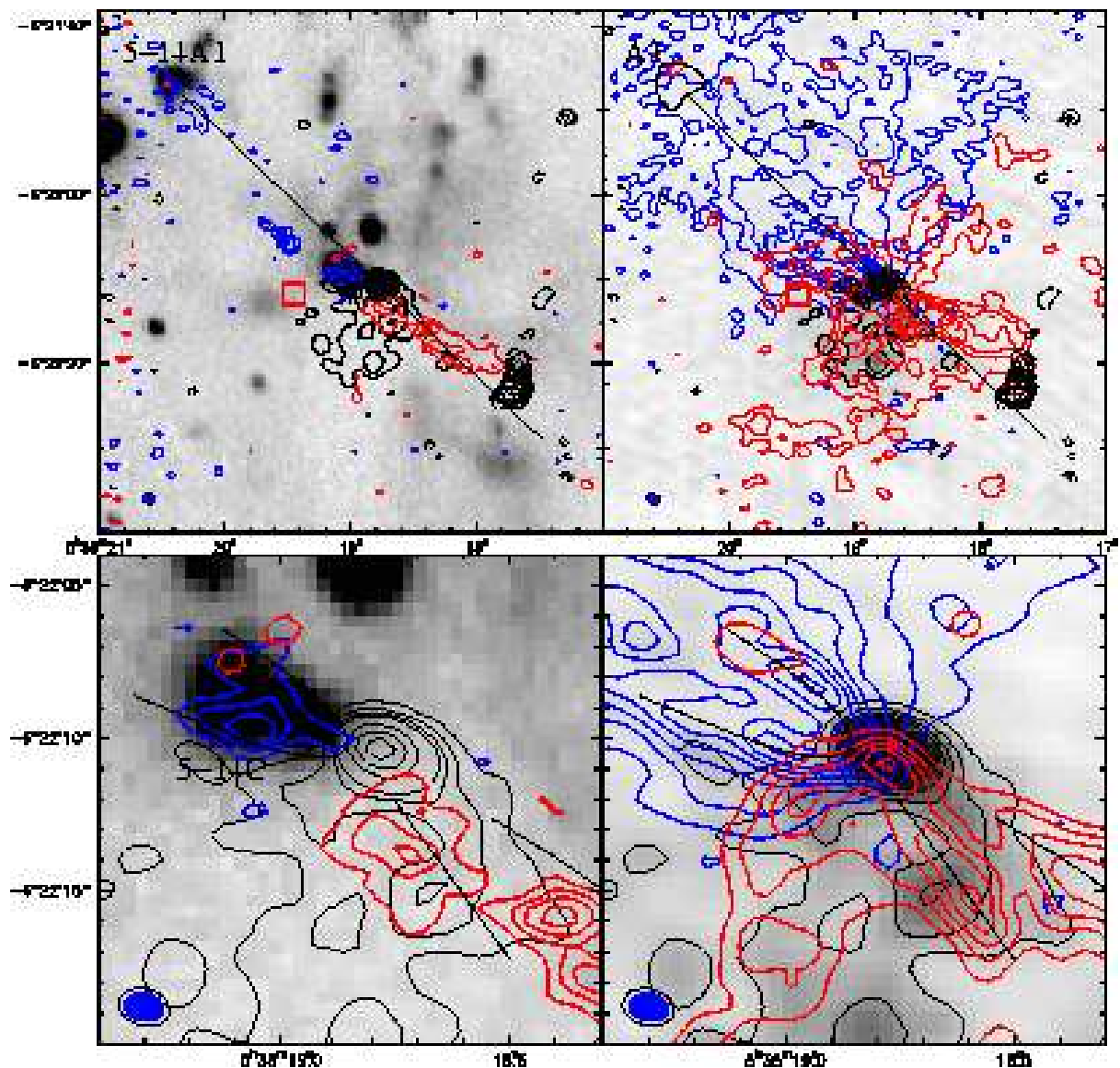}
  \caption{CO(2-1) emission from L1641-N. Blue/red contours:
      blue/redshifted CO; dotted black contours: CH$_3$OH. The square marks
      the position of the AN04 ISO source 18.
      The black/blue ellipses indicate the continuum/CO-line beam size.
      Left: greyscale: 2.12$\mu$m narrow band image (H$_2$ emission +
      continuum). Black contours:  SMA + MAMBO 1.3~mm dust continuum 
      (see Fig.\ \ref{fig_continuum}). The solid line connects the H$_2$
      bow shock A1 and the driving source MM1. The orientation of the CO
      outflows clearly deviates from that line.
      Right: low-velocity CO; greyscale, solid black contours:
      SMA + MAMBO 1.3~mm dust continuum.
      The panels at the bottom show a close-up of
      the central region. Black lines are drawn to emphasize the cavity
      structures seen at low velocities.}
\label{fig_CO_SMA}
\end{figure}

Figure \ref{fig_CO_SMA} shows the spatial distribution of CO line-wing
emission over the field observed with the SMA. The emission has been
split up into a low- and high-velocity component (left/right panel,
respectively). At high velocities, there is evidence for the
presence of a well collimated (though faint), NE-SW oriented molecular
jet. Apparently, it is not centered on the bright MM1 continuum
source, but on MM3. The red-shifted lobe of this collimated flow is
more pronounced and seen over a larger distance from the driving
source. It appears to terminate in a shock feature seen in various other
molecular lines (SO and CH$_3$OH, see below; CS and SiO 
(\cite{chenetal1996}; \cite{mcmullinetal1994}).
The blue-shifted lobe appears to be more spurious. Some blueshifted
high-velocity emission seems to be present also towards the prominent
H$_2$ bow shock A1 seen in the NE corner of Fig.\ \ref{fig_CO_SMA}.

At lower velocities, CO emission is spread over a wide
area. Blue-shifted emission is seen virtually over the entire north-eastern
quadrant, while red-shifted CO prevails towards the south and west of the
cluster center. Centered on MM1 there appears to be a narrow, hollow
cavity-like structure, particularly clear in the blueshifted lobe.
The NE continuum ridge is found just south of the southern edge of the
blueshifted CO cavity, apparently outlining the wall of the outflow cavity.
The small H$_2$ jet formed by the knots labeled as C in Fig.\
\ref{fig_continuum} appears to run right through the CO cavity; however,
it cannot be excluded that this is due to residual error in the astrometry
for the H$_2$ image. 

The data suggest that there are two almost parallel (in projection)
outflows in the cluster center. One is marked by the bipolar,
high-velocity CO jet and apparently driven by MM3. The other one
is indicated by the presence of the bipolar cavity centered on MM1.

It is straightforward to relate the single-dish B-NE lobe and its short,
red-shifted counterlobe to the flows seen at high-resolution with the
SMA. The H$_2$ bow-shock A1 coincides with the point out to which
high-velocity blue-shifted CO is seen in the single-dish data; however, 
at lower velocities, this collimated outflow lobe extends further out
to the north-east. Remarkably, the axis of none of the two flows seen at
high resolution with the SMA really points in the direction of the
A1 H$_2$ bow shock. We suggest that it is MM1, which creates the A1 bow shock
and the bulk of the single-dish B-NE lobe, simply because it seems that
MM1 is the more massive object.

The molecular outflow from L1641-N has generally been interpreted as
a single, wide-angle flow. Our SMA and IRAM 30~m CO observations show
that this is not true. There is clearly outflow along a south-west to
north-east axis,
marked by the flows seen at high-resolution with the SMA and the B-NE
lobe. In contrast, the R-S CO lobe as well as near-infrared and
optical shocks indicate outflow along a N-S axis. The coexistence
of outflow features along both axes within the central arcminute around the
cluster center (CO lobes, optical and infrared H$_2$ shocks) rule out
the possibility that an initially N-S oriented flow could have changed its
orientation to a NE-SW orientation, e.g., as a result of an interaction
of its driving source in a close encounter with another object.
Close inspection of the 30~m CO channel maps also indicates that the N-S
outflow axis, defined by B-N and R-S, is shifted by several arcseconds
to the east of MM1/MM3. We therefore conclude that there are indeed 3 CO
outflows from the cluster center; two, driven by MM1 and MM3, seen with the
SMA along a NE-SW orientation, and a N-S oriented giant outflow. As suggested
already by \cite{stankeetal1998} the dominant mid-infrared source in
the cluster as seen in ground-based and ISO observations (AN04 source 18),
located about
9\arcsec{} to the east of MM1/MM3, appears to drive this outflow. 

\subsubsection{Other lines}

\begin{figure}
  \includegraphics[width=16cm]{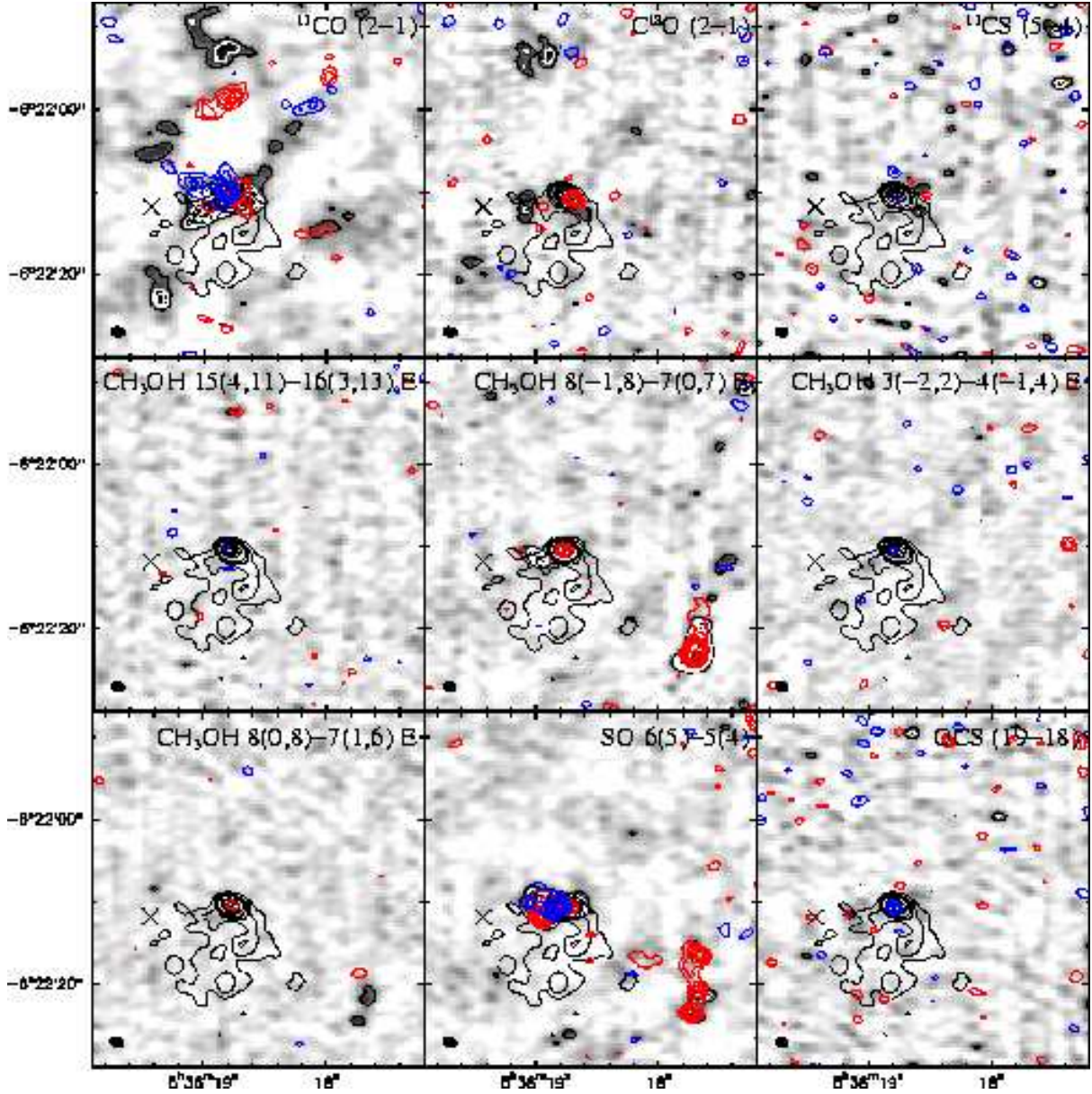}
  \caption{Distribution of molecular line emission in L1641-N. 
  The greyscale and solid black/white contours show line emission
  integrated from 5 to 8.6~km~s$^{-1}$, the
  red and blue contours show emission integrated from 8.6 to 13.4~km~s$^{-1}$
  and from 0.2 to 5~km~s$^{-1}$, respectively.
  The dotted contours show the SMA+MAMBO combined
  continuum map. The $\times$ symbol marks the 
  ISO 7~$\mu$m peak position.
  }
\label{fig_linemaps}
\end{figure}

The SMA provides a frequency coverage of 2~GHz in each of the upper and
lower sidebands and allows several lines to be observed simultaneosly.
Centering the CO(2-1) line in the USB yields the isotopomers $^{13}$CO and
C$^{18}$O in the LSB. In addition to the CO lines, we find some CH$_3$OH
transitions, $^{13}$CS, OCS, SO, and the first few transitions of the
CH$_3$CN~(12-11) series. Fig.\ \ref{fig_linemaps} shows a compilation of
spectral line maps, with emission close to the rest velocity 
(5-8.6~km~s$^{-1}$) and line-wing emission (8.6 to 13.4~km~s$^{-1}$: red;
0.2 to 5~km~s$^{-1}$: blue) plotted separately, and Tab.\ \ref{tab_lines}
gives a list of detected lines.
Besides the lines shown in Fig.\ \ref{fig_linemaps}
there is a marginal detection of the 
HNCO 10(0,10)-9(0,9) transition at 219.798~GHz, which however is too faint for
mapping and will not be discussed further.

$^{13}$CO and C$^{18}$O trace a more
extended structure (the former clearly affected by missing
short spacing information), with some emission associated with MM1 and MM3.
The brightest peak in $^{13}$CO and C$^{18}$O is significantly offset
towards the SW of MM1. MM3 is coincident with the second brightest
$^{13}$CO peak and a faint C$^{18}$O peak; the second brightest
C$^{18}$O peak is seen towards the NE continuum emission ridge. 
Another C$^{18}$O peak is found some 17\arcsec{} north of MM1, roughly
coinciding with the CS peak 3 found by
\cite{chenetal1996}. Besides the CS peak 1, none of the other
CS peaks found by
\cite{chenetal1996} is seen in $^{13}$CO or C$^{18}$O.
$^{13}$CO shows clear signs for emission in the line-wing bins, in a bipolar
fashion around MM1; particularly the blue-shifted emission corresponds well
to the structure seen at the base of the cavity seen in $^{12}$CO 
at moderate velocities. No $^{13}$CO line-wing emission is obviously related
to MM3, in agreement with our suggestion that MM1 is the more massive outflow.

The other molecules show pronounced differences in their distribution.
SO emission is found towards MM1, MM3, and an extended
feature to the SW of the cluster center. Besides two compact features
corresponding to MM1 and MM3 there is also some more extended emission
surrounding these sources, which is seen in the central velocity bin as
well as in the line-wing bins, mostly to the east of MM1. In the integrated
spectrum the blue-shifted wing of the SO line is much more pronounced than the
red wing, whereas in the map, the red wing seems to dominate, particularly due
to strong emission seen towards the SW shock; some more red-shifted SO
emission is seen between the cluster center and the SW shock. 
The SO channel maps (not shown here) suggest that most of the blue-shifted
SO emission is in extended, low surface brightness emission towards
the blue-shifted lobe of the flow from MM1, which is too faint to show up in
the map in Fig.\ \ref{fig_linemaps}.

CH$_3$OH emission is seen mainly towards MM1, and towards MM3 in the 
8($-$1,8)--7(0,7)~E line (and maybe marginally in the 8(0,8)--7(1,6)~E line).
Around MM1 and MM3 the emission in the 8($-$1,8)--7(0,7)~E line appears to be
somewhat extended.
The SW shock is seen in the 8($-$1,8)--7(0,7)~E and 8(0,8)--7(1,6)~E lines,
with a clear redshifted component in the 8($-$1,8)--7(0,7)~E line; it 
extends out to velocities around 15-16~km/s, consistent with the results of 
\cite{mcmullinetal1994} and \cite{chenetal1996} (CS peak 2).
The CH$_3$OH~15(4,11)-16(3,13) line is seen as a faint, compact
feature towards MM1 only; the 3($-$2,2)-4($-$1,4) line in addition
is marginally seen in MM3.

\begin{figure}
  \includegraphics[width=16cm]{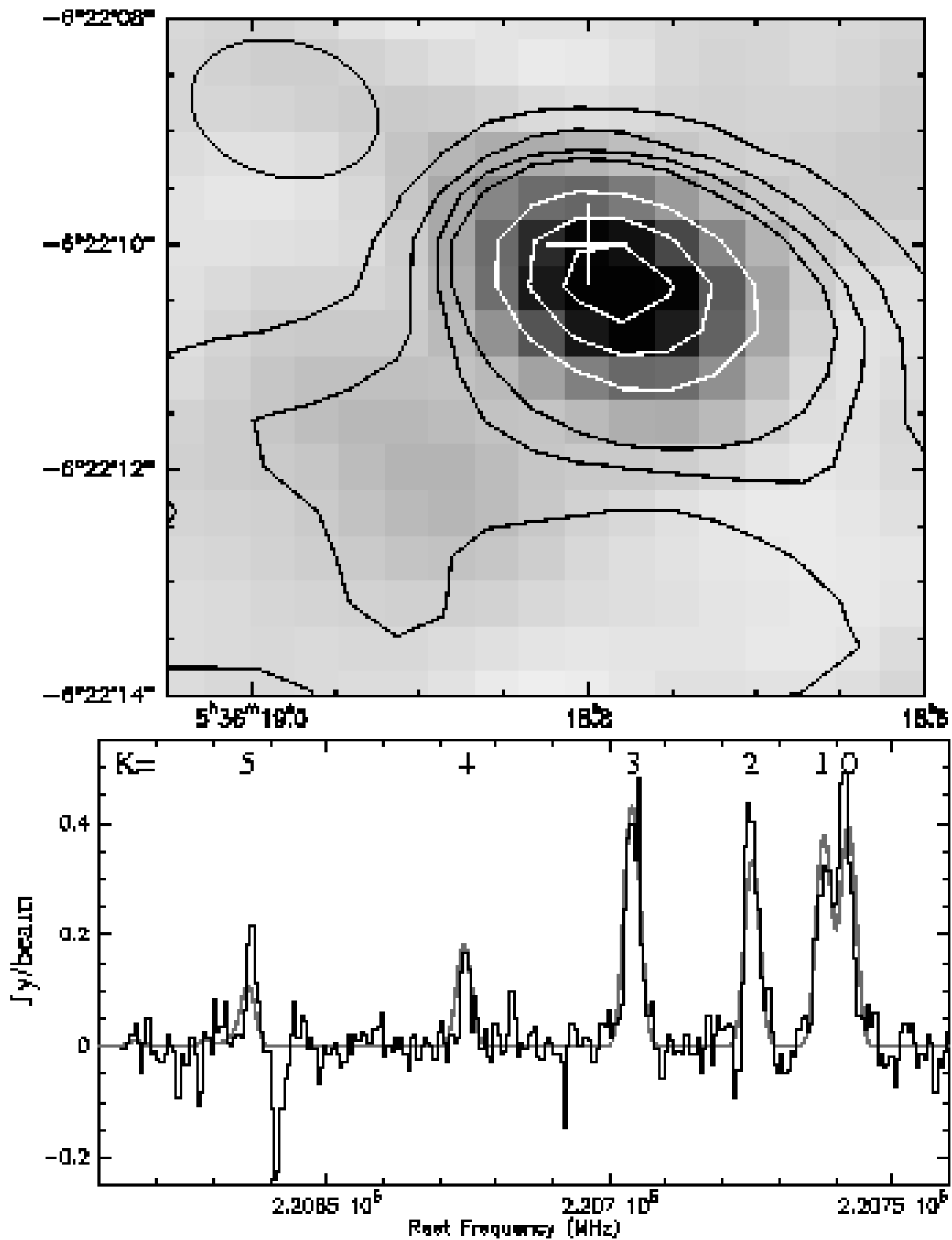}
  \caption{Top: CH$_3$CN map (greyscale; integrated over the $K=0$ and 1
   lines; contours: SMA+MAMBO combined continuum map; cross: phase center;
   ellipse: beam).
   Bottom: CH$_3$CN spectrum of MM1, along with model spectrum (grey line;
   model parameters: $T=130$K, $N_{\rm CH_3CN}=4\cdot10^{15}$cm$^{-2}$, 
   $\Delta v = 4.1$km\,s$^{-1}$, source size 0\farcs42). The dip in the
   spectrum close to the $K=5$
   line is caused by a number of particularly noisy channels.
  }
\label{fig_ch3cn_fit}
\end{figure}

$^{13}$CS appears to be spread out over a similar
area as C$^{18}$O, but is only seen very faintly in the map.
OCS is only seen as a compact feature coincident with MM1.
CH$_3$CN is clearly detected as a compact feature coincident with MM1
and marginally towards MM3 (Fig.\ \ref{fig_ch3cn_fit}, top).

The SO, CH$_3$OH~8($-$1,8)-7(0,7), CH$_3$OH~8(0,8)-7(1,6), and $^{13}$CO line
show significant line wing emission in spectra integrated over the area
covered by the SMA primary beam; faint wing emission might also be seen in
C$^{18}$O. The wing emission is more pronounced on the
blue-shifted side, but fainter red-shifted emission is also present.
Single dish C$^{18}$O, SO and CH$_3$OH spectra (CSO, $\sim$30\arcsec{}
beam, \cite{mcmullinetal1994}) tend to show more
prominent emission in the red wing, suggesting that there is a
significant extended component which is resolved out by the interferometer.
The remaining lines (CH$_3$OH, CH$_3$CN, $^{13}$CS, OCS) as well as
the cores of the lines associated with wing emission are resolved,
with FWHMs around 3~km/s.

MM2 and MM4 are not seen in any of the lines.

Complex organic molecules are thought to form in the dense, dusty
environments around young protostars as the products of grain surface
chemistry evaporate and react with each other \citep{bottinellietal2007}.
Whereas  CH$_3$OH
is clearly associated also with the outflows, this seems not to be the
case for CH$_3$CN and OCS in L1641-N. We have used the XCLASS extension to the
CLASS software (P. Schilke, pers.\ comm.) to simulate the
CH$_3$CN spectrum to estimate the temperature, column density, and
size of the emitting region (Fig.\ \ref{fig_ch3cn_fit}, bottom; we used
a Jy to K conversion factor of 12, corresponding to an observed convolved
source
size of 1\farcs4, i.e., unresolved by our beam; this is a valid assumption,
as the line fluxes integrated over the source are found to correspond well
to the peak fluxes per beam).
The detection of the $K=4$ and 5 lines ($E_{\rm u} = 183$ and 247~K) suggest
a temperature of more than 100~K, and their brightness with respect to the
lower $K$ lines limit the temperature to less than 200~K. The $K=0,1,2$, and 3
lines are of similar brightness, which suggests that they are somewhat
optically thick (for the adopted temperature range, the $K=3$ line ($E_{\rm u}
= 133$~K) should be
clearly brighter than the $K=2$ line if optically thin, due to its high
statistical weight). The observed line intensities then imply a 
CH$_3$CN column density of the order of a few times 10$^{15}$cm$^{-2}$ and a
source size of the order of 0\farcs4, i.e., 150-200~AU. This is significantly
smaller and somewhat cooler than CH$_3$CN emission regions found in hot cores
in massive star forming regions (e.g., W3(H$_2$O), \cite{chenetal2006})
but is is comparable to the hot 'corinos' which have now been found
in a handful of other low-mass protostars 
\citep[e.g.,][]{bottinellietal2004,joergensenetal2005}.

\subsubsection{Evidence for a rotating disk in MM1}

\begin{figure}
  \includegraphics[width=10cm]{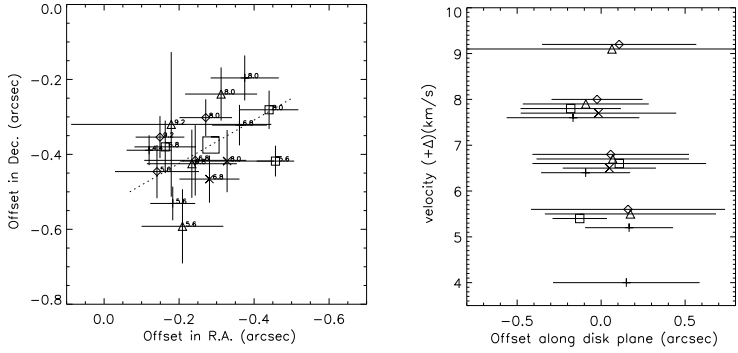}
  \caption{Left: Position of the emission maxima in individual
  velocity channels for
  the CH$_3$CN (diamonds: K=0; triangles: K=1; squares: K=2; x's: K=3)
  and OCS lines ($+$); the size of the crosses show the 1-$\sigma$
  positional errors as given by the gaussian fitting procedure.
  Offsets are given in arcseconds from the phase center. The square marks
  the position of the MM1 continuum peak, the dotted line the orientation
  of the assumed disk plane (position angle 120$^\circ$).
  Right: velocity vs.\ offset from source along the disk plane.
  The plotting symbols are the same as in the left panel. For clarity,
  velocities in the various transitions are plotted with a small offset.
  Although the positional errors are larger than the positional offsets from
  channel to channel, a trend for increasing velocities (going from
  southeast to northwest) is consistently seen in all transitions.
  }
\label{fig_veloffsets}
\end{figure}

In Fig.\ \ref{fig_veloffsets} (left panel) the positions of the emission
maxima, as derived
from gaussian fits, are plotted for individual 1.2~km~s$^{-1}$ wide channels
for the CH$_3$CN K=0, 1, 2, and 3 lines and the OCS line. Only channels in
which emission above 5$\sigma$
is seen are used. We chose to use only CH$_3$CN and OCS lines in this plot as
it is these molecules where we see the least contamination by outflow related
emission. 

Overall, the positions of the emission maxima define a south-east to 
north-west oriented, elongated structure (position angle $\sim120^\circ$).
This is unlikely to be the result
of positional measurement errors, as this would tend to produce a distribution
which is elongated along the beams major axis, which is at a position angle of
about 75$^\circ$. 
There is a clear trend for the lower velocity center positions to cluster to
the south-east of the MM1 central position, and the higher velocity positions
to be found to its north-west. To illustrate this further, the right panel
of Fig.\ \ref{fig_veloffsets} shows the channel velocities as a function of
the offset along the disk plane. Although the positional errors are larger
than the offsets, a trend for velocities increasing from sout-east to
north-west is consistently seen in all transitions. Clearly, better data
are needed to confirm this observation.

We tentatively suggest that the positional arrangement of the emission maxima 
indicates a flattened structure around the protostellar source embedded in
MM1; a similar result has been obtained by \cite{cesaronietal1997,cesaronietal1999} for the intermediate-mass protostar IRAS\,20126$+$4104, and has been interpreted as evidence for a circumstellar disk. The velocity gradient hints at rotation, with a rotation axis being 
roughly parallel to the axis of the outflow from MM1. This would imply that the 
CH$_3$CN and OCS emission arises from grain mantle evaporation in
a heated disk rather than the dense, infalling envelope. However, higher
angular resolution observations resolving the source is required
to discriminate between these scenarios (is it spherical or flattened?)
definitively.
Also, neither
the angular resolution nor the velocity resolution nor the sensitivity of
our data are good enough to derive estimates of the central object's mass or
the size of the disk.

\section{Summary and Discussion
\label{chap_summary}}

We have presented new 1.3~mm dust continuum and molecular line observations
of the L1641-N cluster area. Single-dish wide field CO mapping revealed the
presence of several blue- and red-shifted outflow lobes, which can be grouped
into at least 4 bipolar flows: a north-south oriented giant outflow B-N/R-S,
driven most likely by the AN04 ISO source 18; a
shorter north-east to south-west oriented outflow B-NE/R-W driven by the
dominant millimeter continuum source in the cluster center; another parsec
scale outflow B-E/R-E/R-SW corresponding to the H$_2$ flow 51 of
SMZ2002, driven by
an embedded ISO and millimeter source east of the cluster center; and a
small flow B-SE/R-SE driven by a near-infrared star south-east of the cluster
center. Moreover, some CO outflow activity might also be associated with
the Strom~11 group further south-east of the L1641-N cluster center (B-SE2).

The SMA observations show that there are in fact two deeply embedded sources
in the very cluster center, MM1 and MM3, each of which drives a CO outflow,
which together make up the B-NE/R-W outflow:
MM1 is at the center of two cavity-shaped outflow lobes seen at moderate CO
velocities, and MM3 drives a collimated, faint jet seen only at high
velocities. The flow from MM1 carries more mass as only this flow is
seen in the optically thin $^{13}$CO. We find $^{13}$CO, C$^{18}$O, and 
$^{13}$CS emission tracing the core from which MM1 and MM3 are forming,
CH$_3$OH and SO emission tracing the protostellar sources MM1 and
MM3  as well as their outflows, and OCS and CH$_3$CN emission from MM1.
Analysis of the CH$_3$CN $J=12-11$ ($K=0...5$)
lines suggests that the emitting gas traces a compact (100-200~AU),
warm (100-200~K) region deep inside the MM1 envelope. This could be the warm,
innermost portion of the infalling envelope, i.e., a hot corino,
as seen in other low-mass protostars. Alternatively, it may be the
the heated surface layer of a disk, as suggested by our tentative
detection of a velocity gradient over the CH$_3$CN source perpendicular to the
MM1 outflow axis.

Widespread CO line-wing emission is seen virtually over the entire cluster
core area. In total, we estimate
that a few solar masses of molecular gas have been set into motion. This
estimate does not include corrections for optical depth effects or outflow
gas having low radial velocities, however, so the total outflow mass might
well be larger by a factor of $\sim 10$. The flows are estimated to carry 
$\sim$16~M$_\odot$~km~s$^{-1}$ of momentum and a kinetic energy of 
75~M$_\odot$~km$^2$s$^{-2}$, which are lower limits again; correcting the mass
for optical depth and for CO with v$_{\rm rad}$ about 0, and correcting the
velocities for inclination might well increase these values by more than an
order of magnitude. 

In clusters of low mass stars, protostellar outflows can impart a
significant amount of energy and momentum into their surroundings and,
perhaps, thereby regulate the star formation rate 
\citep{normansilk1980,matznermckee2000,linakamura2006}.
They may also play an important role in
replenishing turbulence, which hydrodynamic as well as magnetohydrodynamic
simulations suggest decays on timescales comparable to
a clouds free-fall timescale $\tau_{\rm ff} = (3\pi/32G\rho)^{1/2}$
\citep[e.g.][]{maclow1999}. In order to maintain turbulence, kinetic energy
and momentum has to be provided by the flow at a high enough rate to
compensate for the loss due to the decay of turbulence. 

The mass of the L1641-N cluster forming core
is a few times 100~M$_\odot$ \citep{reipurthetal1998,tatematsuetal1993},
a certain part of which is in the still inactive 'tail' extending
south-east of L1641-N/Strom~11; the actual mass involved in directly building
the L1641-N cluster might be about 100-200~M$_\odot$. Using a cloud
radius of about 0.5~pc (estimated from the map shown by 
\citet{reipurthetal1998}) and a mass $M_{core}$ of 200~$M_\odot$ yields
$\tau_{\rm ff} = 4.1\cdot 10^5$yr.
We estimate a one-dimensional velocity dispersion $\sigma_{\rm 1D}$
of 0.94~km~s$^{-1}$ for
the L1641-N cloud core based on the addition (in quadrature) of the intrinsic
core dispersions of 0.72~km~s$^{-1}$ and core-core motion, 0.6~km~s$^{-1}$
in \citet{tatematsuetal1993}; cores 67 and 69).
The rate $L_{\rm turb} = E_{\rm turb}/\tau_{\rm ff}$
at which kinetic energy is lost due to decaying turbulence is then 
estimated to be
0.1~L$_\odot$ (where $E_{\rm turb} = 3/2 M_{\rm core}\sigma_{\rm 1D}^2$).


The momentum available in the CO flows has the potential to add significant
motions to the cloud core. If the forward momentum were completely transferred
to the 200~M$_\odot$ of core gas, it would accelerate it by at least
$\sim$0.1~km~s$^{-1}$ and possibly up to the cloud velocity dispersion
with the corrections mentioned above.
In addition, sideways motions can be excited, which does not affect
the flows forward oriented momentum. In terms of energy supply, even
the (lower limit on) $L_{\rm mech}$ of 1.21~$L_\odot$ is greater than 
$L_{\rm turb}$. However, a large fraction of that can be expected to be
dissipated in shocks.

The rate $L_{\rm gain}$ at which the cloud gas gains energy, taking into
account radiative losses, can be estimated as follows:
we assume that all of the outflow forward momentum is dumped
in the clump and used to drive the turbulence,
and that the outflows accelerate a fraction
$M_{\rm c}$ of the clump's mass to a velocity equal to the clump's
three-dimensional velocity dispersion $\sigma_{\rm 3D}$. 
The clump mass set in motion per unit time is then 
$\dot{M}_{\rm c}=F_{\rm out}/\sigma_{\rm 3D}$
where $F_{\rm out}$ is the sum of outflow momentum input rates.
Therefore,
$$L_{\rm gain}
  =\frac{1}{2}\dot{M}_{\rm c}\sigma_{\rm3D}^2=\frac{\sqrt{3}}{2}F_{\rm out}\sigma_{\rm 1D}$$
where $\sigma_{\rm 1D}$ is the observed, one-dimensional velocity dispersion.
With $\sigma_{\rm 1D}$ of 0.94~km\,s$^{-1}$ and the sum of outflow momentum rates
of $F_{\rm out}=13.6\times 10^{-4}~M_\odot$~km\,s$^{-1}$\,yr$^{-1}$ from
Table 1 the energy gain rate is $L_{\rm gain} \sim 0.18 L_\odot$.
This result is comparable to the estimate of the clouds rate of
energy loss $L_{\rm turb} \sim 0.1 L_\odot$, particularly if
keeping in mind that the outflow mass and velocity measurements are
lower limits due to optical depth and projection, respectively.
The above calculation indicates that the molecular outflows can
indeed provide sufficient energy and momentum
to sustain cloud turbulence.

Protostellar outflows are known to be collimated.
In regions forming only a small number of isolated and well separated
protostars, the outflow will only affect a small fraction of the total
cloud volume and their impact is therefore relatively small.
However, in regions of clustered star formation, the protostellar density
is higher and a larger
fraction of the cloud volume will be affected. We have found wide-spread
CO line-wing emission in L1641-N, indicating that the presence of several
CO outflows does affect a significant volume of the cluster core. In NGC~1333,
\cite{quillenetal2005} found a system of cavities permeating the cloud
and suggested, if these are relic outflow events, then $\sim 10$\% of
the cloud was affected.
\cite{williamsetal2003} found that
most of the CO line wing emission in the OMC-2 and OMC-3 star forming regions
is in broadly dispersed, extended features, which also suggests that a large
fraction of the cloud volume is affected by outflows.
From the small fraction of the high velocity emission in young, compact
outflows, and using a calculation similar to the above,
they also estimated that the energy injection rate was sufficient
to maintain cloud turbulence.

L1641-N, NGC~1333, and OMC-2/3 are regions forming modest sized clusters of
a few tens to hundreds of stars. In the case of L1641-N, \cite{hodappdeane1993}
showed that star formation has been going on for about 10$^6$~years at
2--3 stars every 10$^5$ years. Assuming that protostars
drive powerful CO outflows over the first 1--2$\times$10$^5$ years of their
evolution, the 5 CO outflows we found in the region suggest that star formation
is continuing at a similar pace. NGC~1333 also shows evidence for star
formation over million year time scales \citep{ladaetal1996}, 
which is still going on, and even the IC~348 cluster might still
be in the buildup phase \citep{eisloeffeletal2003}. These modest sized
cluster forming regions
typically feature a handful to one to two dozens of energetic outflows.
Assuming that the formation of a more massive
cluster (few thousand or ten thousand stars) also proceeds more or less
steadily over a few million years rather than in one short burst, then one
can expect that at any time there are many tens or hundreds of stars in 
their youngest evolutionary stage and driving very energetic outflows.
It can then be expected that virtually the entire cluster volume is affected
by outflow activity.


The action of
a population of outflows from low-mass protostars might also have consequences
for high-mass star formation. 
In the `turbulent core' scenario 
\citep[e.g.,][]{mckeetan2003}, massive stars form from high density cores
formed within a high pressure, turbulent clump. Accretion rates are high
enough to overcome the radiation pressure of the central object. 
The formation time scale $t_{*f}$ is comparable to the free fall timescale
of the surrounding clump, i.e., to the timescale of turbulence decay. Hence
it might not really be crucial to maintain the turbulence {\em while} the
massive star forms. But it might be a prerequisite, to keep the turbulence
in the high pressure environment {\em until} the
massive star forming core has formed and reached the point when it starts to
collapse.

The `competitive accretion' scenario \citep[e.g.,][]{bonnelletal2001}
requires that gas falls into the central region of a forming cluster, allowing
the protostars to grow beyond the masses of their original envelopes. This
infall might be counteracted by outflows from the protostars in the cluster
center. Furthermore, \citet{krumholzetal2005} claim that competitive accretion
and collisional growth of protostars \citep{bonnelletal1998} requires
efficient removal of kinetic energy from the cluster forming gas, which
could also be prevented by low-mass protostars outflows. 

\acknowledgments

We thank Jan Forbrich and the IRAM staff for their help with the 30~m
observations, and Babar Ali for providing fits files of his ISO maps.
We are thankful to Peter Schilke for providing and helping
with the XCLASS software. We thank the referee for a very careful reading of
the manuscript. The Submillimeter Array is a joint project between
the Smithsonian Astrophysical Observatory and the Academia Sinica Institute
of Astronomy and Astrophysics and is funded by the Smithsonian Institution
and the Academia Sinica.
This work was supported by the Alexander von Humboldt Stiftung,
and NSF grant AST-0324328.




\clearpage

\begin{deluxetable}{rccccccccc}
\tabletypesize{\scriptsize}
\tablewidth{0pt}
\tablecaption{Parameters of the individual CO outflow lobes. Momentum and
kinetic energy have been calculated assuming $v_{\rm cen} = 6.8$~km/s.
In the last two rows, mass, $E_{\rm kin}$, and $P$ are integrated over the
entire area where high velocity CO emission is seen, and includes a
diffuse component which can not be attributed to individual flows;
however, the velocity range (close to the ambient cloud
velocity) had to be chosen more carefully than for the individual CO flow
lobes, therefore omitting some of the slow moving line-wing emission.
$L_{\rm mech}$ and $F_{\rm mech}$ in the last two rows are given as
the sum over the individual outflow lobes.
\label{tab_COlobes}}
\tablehead{
\colhead{} & \colhead{$\Delta$v} & \colhead{Mass} & 
\colhead{$E_{\rm kin}$} & \colhead{$P$} & \colhead{$v_{\rm max}$} & \colhead{$l$}&
\colhead{$\tau$} & \colhead{$L_{\rm mech}$} & \colhead{$F_{\rm mech}$}\\
\colhead{} & \colhead{km~s$^{-1}$} & \colhead{$M_\odot$} & 
\colhead{$M_\odot$ km$^2$~s$^{-2}$} & \colhead{$M_\odot$ km~s$^{-1}$} &
\colhead{km~s$^{-1}$} & \colhead{pc} & \colhead{10$^3$~yr} & 
\colhead{$L_\odot$} & 10$^{-4} M_\odot$km s$^{-1}$yr$^{-1}$
}
\startdata
B-N \& B-NE & $-$29.8 \ldots  3.8 &  0.7 &  28  & 4.9 & -28.8 & 0.21& 7.1 &0.65 &6.9\\
R-W         & 11.0 \ldots 43.4    & 0.07 & 6.1  & 0.8 & 30.0  & 0.23& 7.5 &0.13 &1.0\\
B-E         & $-$5.8 \ldots 6.2   & 0.27 & 1.8  & 0.8 & -10.8 & 0.13& 11.8&0.025&0.7\\
R-E         & 11.0 \ldots 33.8    & 0.07 & 4.4  & 0.7 & 24.0  & 0.09& 3.7 &0.20 &1.9\\
R-SE        & 9.8    \ldots 26.6  & 0.036& 1.4  & 0.29& 14.4  & 0.04& 2.7 &0.09 &1.1\\
B-SE        & $-$2.2 \ldots 6.2   & 0.044& 0.18 & 0.11& -4.8  & 0.05& 10.2&0.003&0.1\\
B-SE2       & $-$2.2 \ldots 5.0   & 0.13 & 0.68 & 0.38& -7.2  & 0.18& 24.5&0.005&0.16\\
R-SW        & 9.8 \ldots 21.8     & 0.34 & 6.21 & 2.0 & 12.0  & 0.65& 53.0&0.02 &0.37\\
R-S         & 9.8 \ldots 26.6     & 1.1  & 33   & 7.6 & 14.4  & 0.86& 58.5&0.09 &1.3\\
blue total  & $-$29.8 \ldots 3.8  & 0.95 & 27   &  5.8&       &     &     &0.68 &7.86\\
red total   & 12.2 \ldots 39.8    & 1.24 & 48   & 10.2&       &     &     &0.53 &5.67\\
\enddata
\end{deluxetable}

\begin{table*}
  \caption{Molecular lines detected, and the features where the lines are seen
   ($\times$ means detected, $(\times)$ means marginally detected, and -- means
   not detected).
    }
  \label{tab_lines}
  \begin{tabular}{ccllcccc}
    \hline
        $\nu_{\rm rest}$ (GHz) & $\nu_{\rm obs}$ (GHz) &Species&Transition & MM1 & SW shock & outflow & MM3\\
    \hline
219.5604 & 219.5601 & C$^{18}$O & 2-1&$\times$ & -- & -- & $\times$\\
219.9494 & 219.9497 & SO        & 6(5)-5(4)&$\times$&$\times$&$\times$&$\times$\\
220.0785 & 220.0789 & CH$_3$OH  & 8(0,8)-7(1,6) E&$\times$&$\times$&--&($\times$)\\
220.3987 & 220.3994 & $^{13}$CO & 2-1 &$\times$ & -- &$\times$&$\times$\\
220.6411 & 220.6424 & CH$_3$CN  & 12(5)-11(5)& $\times$ & -- & -- & --\\
220.7090 & 220.7098 & CH$_3$CN  & 12(3)-11(3)& $\times$ & -- & -- &($\times$)\\
220.7303 & 220.7307 & CH$_3$CN  & 12(2)-11(2)& $\times$ & -- & -- &($\times$)\\
220.7430 & 220.7432 & CH$_3$CN  & 12(1)-11(1)& $\times$ & -- & -- &($\times$)\\
220.7473 & 220.7469 & CH$_3$CN  & 12(0)-11(0)& $\times$ & -- & -- &($\times$)\\
229.5891 & 229.5875 & CH$_3$OH  & 15(4,11)-16(3,13) E & $\times$ & -- & -- & --\\
229.7588 & 229.7574 & CH$_3$OH  & 8($-$1,8)-7(0,7) E & $\times$ & $\times$ & -- & $\times$ \\
230.0270 & 230.0261 & CH$_3$OH  & 3($-$2,2)-4($-$1,4) E & $\times$ & -- & -- & -- \\
230.5380 & 230.518  & CO & 2-1 \\
231.0610 & 231.0601 & OCS       & 19-18 & $\times$ & -- & -- & -- \\
231.2208 & 231.2185 & $^{13}$CS & 5-4 & ($\times$) & -- & -- & -- \\
\multicolumn{4}{l}{continuum (mJy)} & 470 & & & 79\\
\hline
  \end{tabular}
\end{table*}

\end{document}